ORIGINAL ARTICLE

# Controlling light with light using coherent metadevices: all-optical transistor, summator and invertor

Xu Fang[1], Kevin F MacDonald[1] and Nikolay I Zheludev[1,2]

Although vast amounts of information are conveyed by photons in optical fibers, the majority of data processing is performed electronically, creating the infamous 'information bottleneck' and consuming energy at an increasingly unsustainable rate. The potential for photonic devices to directly manipulate light remains unfulfilled due largely to a lack of materials with strong, fast optical nonlinearities. In this paper, we show that small-signal amplifier, summator and invertor functions for optical signals may be realized using a four-port device that exploits the coherent interaction of beams on a planar plasmonic metamaterial, assuming no intrinsic nonlinearity. The redistribution of energy among ports can provide nonlinear input-output signal dependencies and may be coherently controlled at very low intensity levels, with multi-THz bandwidth and without introducing signal distortion, thereby presenting powerful opportunities for novel optical data processing architectures, complexity oracles and the locally coherent networks that are becoming part of the mainstream telecommunications agenda.

## INTRODUCTION

Photonic technologies are central to our information-based society: optical fibers deliver telephone and internet traffic around the globe, while semiconductor lasers and photodiodes are employed to generate and receive data. However, with data traffic increasing at a rate of 40%–50% per annum, the fundamental limitations of the current hybrid technology platform (in which information is transported optically but processed and routed in electronic circuits), specifically the electronic bit-rate bottleneck and increasing energy consumption, are of growing significance.[1] Today's networks are largely optically opaque; so conversion to all-optical switching of data will address the former by enabling increased bit rates and low latency movement of data through the fiber infrastructure with greater routing agility and simplicity. In relation to the latter, the power requirements of data centers and global communication systems are rapidly becoming unsustainable. Although optics can fundamentally save energy in interconnects, energy-efficient optical switches, which must be at least as fast as and more efficient than their electronic counterparts, have not yet been developed: a substantial reduction in energy-per-bit performance for all-optical switching to the 10 fJ/bit level[2] is essential.

A few years ago, David Miller presented a well-argued discussion on the requirements for a practical all-optical digital switch that could compete with and surpass electronic devices,[3] concluding with the widely accepted viewpoint that the all-important optical transistor, with properties that are comparable to electronic transistors, does not currently exist. Substantial effort is now focused on finding the fast and highly nonlinear media that would enable such devices. This approach is based on the premise that controlling light signals with light intrinsically requires intense optical fields to facilitate beam interactions in nonlinear media (the Huygens superposition principle dictating that light beams in a linear medium will pass though one another without mutual disturbance). The most promising research directions here include the exploitation of gain nonlinearity in active media,[4,5] carrier-induced nonlinearity in semiconductor photonic crystal cavities,[6] and the enhancement of nonlinear responses in metals, semiconductors and low-dimensional carbon with hybrid metamaterial structures.[7] The latter can provide picosecond to sub-picosecond response times with switching energies of approximately 100 pJ per pulse—still far from the target energy/bit goal. Optomechanical and phase-change metamaterials[8–10] are still evolving and could potentially achieve all-optical switching at the fJ/bit level, but they will likely be constrained to operate with micro- to nanosecond response times that are not sufficiently short for key telecommunications applications.

In recent years, the topic of optical computing has re-emerged, notably in regard to the practicality of optical switching based on so-called 'zero-energy' interference devices[11–13] and related optical logic architectures,[14] as well as concepts for non-Turing optical computing.[15–17] In this paper, we introduce another approach to all-optical data processing using a device that enables the realization of basic small-signal amplification (cf. transistor) and logic (signal summation and inversion) functions. Like the previously mentioned zero-energy devices,[11] this solution implements optical data processing functions without optically nonlinear media and can therefore operate at very low power levels. Their functionality is underpinned by the re-distribution of energy among ports and does not induce harmonic distortion of the information carrier.

Because these devices exploit the coherence of optical beams and their ability to interfere, they may be described as 'coherent control' data processing devices. Coherent control is a well-understood concept in quantum mechanics,[18] where it is used to direct dynamic processes with light by engaging quantum interference phenomena. Coherent control concepts have been employed to manipulate various processes, such as the direction of electron motion in semiconductors,[19] the breaking of chemical bonds[20] and the absorption and localization of light.[21–33] Recent studies have also experimentally demonstrated that the coherent absorption process in an ultrathin metamaterial layer can facilitate light-by-light modulation with multi-THz bandwidth[34–36] (in principle, limited only by the spectral width of the metamaterial absorption resonance) and at the single-photon ($\sim 2 \times 10^{-19}$ J) level,[37] and that polarization and refraction effects in planar metamaterials can be similarly coherently modulated.[38,39] We demonstrate that exploitation of the coherent control paradigm in media of subwavelength thickness may facilitate the optical realization of key data processing components, such as an analog summator, an invertor and a small-signal gain device (cf. transistor), and we illustrate that photonic metamaterials, with properties that can be engineered by design (and encapsulated in a single complex scattering parameter for analytical purposes), can deliver the requisite material parameters to achieve these functions.

## MATERIALS AND METHODS

A generic coherent control device that can operate in multiple modes, for example as an optical gate or amplifier, is a four-port device with two inputs and two outputs. In the simplest case, this device may comprise a thin layer of absorbing material illuminated from both sides by two mutually coherent light waves that represent the two input signals. The transmitted and reflected light waves that propagate in either direction away from the film constitute the output signals. Consider a material layer illuminated at normal incidence by two counter-propagating coherent optical waves $E_\alpha$ and $E_\beta$; we denote the two output waves $E_\gamma$ and $E_\delta$ (Figure 1). In all cases, we assume an absorbing film with linear optical properties, and that the polarization states of the waves do not change as a result of interaction with the film. We also assume that light–matter interactions in the film are of an electric dipole nature, and thus that only the electric field components of the electromagnetic waves are relevant. Under these conditions, the two input fields and two output fields are related by a complex scattering matrix $S$:[40]

$$\begin{bmatrix} E_\delta \\ E_\gamma \end{bmatrix} = \begin{bmatrix} S_{11} & S_{12} \\ S_{21} & S_{22} \end{bmatrix} \begin{bmatrix} E_\alpha \\ E_\beta \end{bmatrix} \quad (1)$$

Any all-optical gating function requires a nonlinear relationship between the intensities of the signal input and signal output waves, and we do not challenge this wisdom. However, while many known gating solutions are based on exploiting the intrinsic optical nonlinearity of a material to achieve this function, we rely in the present case on a nonlinearity derived from the coherent nature of beam interactions and the properties of matrix (1): the linearity of the film and therefore the scattering matrix imply that a proportional scaling of both input signals by a factor $\eta$ will lead to a correspondingly proportional scaling of both output signals:

$$\begin{bmatrix} \eta E_\delta \\ \eta E_\gamma \end{bmatrix} = \begin{bmatrix} S_{11} & S_{12} \\ S_{21} & S_{22} \end{bmatrix} \begin{bmatrix} \eta E_\alpha \\ \eta E_\beta \end{bmatrix} \quad (2)$$

However, this does not imply that increasing one input signal will proportionally increase one or both of the output signals, so it is generally the case that:

$$\begin{bmatrix} \eta E_\delta \\ \eta E_\gamma \end{bmatrix} \neq \begin{bmatrix} S_{11} & S_{12} \\ S_{21} & S_{22} \end{bmatrix} \begin{bmatrix} \eta E_\alpha \\ E_\beta \end{bmatrix} \quad (3)$$

Thus, in a four-port device as described by Equation (1), the relationship between one input port and one output port can, counter-intuitively, be nonlinear. We show how this fact can be exploited to useful effect.

Now consider a film which is sufficiently thin that wave retardation across its thickness can be ignored and each constituent molecule assumed to be exposed to the same electric field, i.e., the combined field of the incident waves $E_\alpha + E_\beta$. The molecules in the film will re-radiate absorbed energy equally in the forward and backward directions with an efficiency that is dependent on the wavelength of excitation $\lambda$ and proportional to the driving field, i.e., proportional to $s(\lambda)(E_\alpha + E_\beta)$, where $s(\lambda)$ is the complex wavelength-dependent amplitude scattering coefficient of the film for a single incident beam.[41] The magnitude of $s(\lambda)$ corresponds to the relative amplitude of the re-radiated field and its phase to the phase lag between the re-radiated and driving fields. It may include losses in the film material, thereby excluding any assumption of equality between the combined incident and combined output intensities. Field continuity subsequently dictates the following exact scattering matrix expression for a coherent four-port device based on a vanishingly thin film:

$$\begin{bmatrix} E_\delta \\ E_\gamma \end{bmatrix} = \begin{bmatrix} s(\lambda) & s(\lambda)+1 \\ s(\lambda)+1 & s(\lambda) \end{bmatrix} \begin{bmatrix} E_\alpha \\ E_\beta \end{bmatrix} \quad (4)$$

which reduces for single-beam illumination, with either of the two inputs set to zero, to stipulate that the reflected and transmitted fields are equal respectively to the re-radiated field and to the sum of the incident and re-radiated fields. This expression can serve as an approximation to Equation (1) for realistic (i.e., finite thickness) materials in cases where the contribution from interference among multiply reflected/transmitted beams is small. We illustrate below that it can reasonably be applied to planar plasmonic metamaterials to streamline the analysis and interpretation of four-port photonic device characteristics.

## RESULTS AND DISCUSSION

We begin analysis of the four-port device by highlighting two important cases: If the incident waves ($\alpha$ and $\beta$) have the same amplitude $|E_\alpha|=|E_\beta|$, they will form a standing wave along the direction normal to the plane of the film. If the thin film is positioned at a node of the standing wave, then $E_\alpha = -E_\beta$. Here, regardless to the value of $s(\lambda)$, $E_\gamma$ will always be equal to $E_\alpha$, and $E_\delta$ equal to $E_\beta$. This corresponds to what is known as 'coherent perfect transmission'—a situation in which light does not interact with the film because it is located at a point where the combined electric field vanishes. In contrast, if the material layer is located at an anti-node of the standing wave, where $E_\alpha = E_\beta$, then both $E_\gamma$ and $E_\delta$ will be zero if $\mathrm{Re}\{s(\lambda)\} = -0.5$ and $\mathrm{Im}\{s(\lambda)\} = 0$ (whereby the film exhibits the maximum possible level of 'zero-thickness' single-beam absorption,[42,43] which is 50%). This phenomenon is known as 'coherent perfect absorption'. (Although perfect transparency is contingent on the subwavelength thickness of an absorber,[34] coherent perfect absorption *per se* is not. The latter has thus been considered in a variety of optically thick material systems and spectral domains.[28,44,45]) In what follows, we also explore intermediate values of the incident waves' mutual phase $\theta = \mathrm{Arg}\{E_\alpha\} - \mathrm{Arg}\{E_\beta\}$ at an ultra-thin material layer.

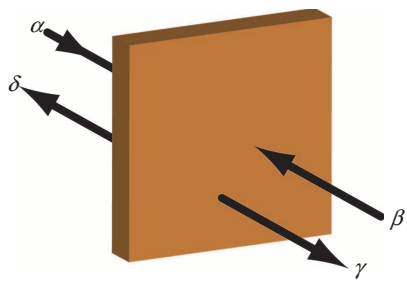

**Figure 1** Generic four-port coherent control data processing device: a layer of material is illuminated from either side by counter-propagating coherent input light waves $\alpha$ and $\beta$; waves $\gamma$ and $\delta$ are the device outputs. The electric fields of the incident waves $E_\alpha$ and $E_\beta$ and the output waves $E_\gamma$ and $E_\delta$ are related by the complex scattering matrix $S$.

As previously discussed, the characteristic nonlinear dependence of the intensity (defined as $I=EE^*$) of a given four-port device output wave on the intensity of a given input is the foundation of the functionalities described in this study. To reemphasize, this must not be confused with the purely linear nature of the film material's optical response, which implies that a proportional simultaneous increase in the intensities of *all* input waves will produce a corresponding proportional increase in the intensities of all output waves. Consider the intensity of output $\gamma$:

$$I_\gamma = |1+s(\lambda)|^2 \, I_\alpha + |s(\lambda)|^2 \, I_\beta \\ + 2\,\mathrm{Re}\left\{(1+s(\lambda))s^* E_\alpha E_\beta^*\right\} \quad (5)$$

If the input intensities $I_\alpha$ and $I_\beta$ are proportionally increased, there will be a correspondingly proportional increase in $I_\gamma$. However, if only $I_\beta$ changes, then $I_\gamma$ will respond in a nonlinear fashion. For example, if $s(\lambda) = -0.5 + 0i$, then with $I_\alpha$ fixed the dependence of $I_\gamma$ on $I_\beta$ is not only nonlinear but also non-monotonous for $\theta < 90^0$, as illustrated in Figure 2 (where $I_\alpha = 1$). If one of the input signals is removed, the device reverts to a truly linear single-beam mode of operation in which the output signals are strictly proportional to the remaining input.

It should be noted that when light propagates through a conventional nonlinear medium, harmonic distortion occurs, which can cause optical instability and multi-stability in extreme cases. The nonlinear behavior of the four-port device considered in this paper is very different: its functionality is underpinned by the re-distribution of energy among different ports and does not cause harmonic distortion of signals. Although counterintuitive in many instances, this redistribution is based only on linear interference and as such is strictly compliant with energy conservation requirements; the enabling characteristic of the four-port coherent device being that the level of absorption is not fixed, but instead is strongly dependent on the mutual intensity and phase of the input beams.

So far, this analysis has been based on an ideal, vanishingly thin absorbing layer that can be described by a single, wavelength-dependent complex scattering parameter $s(\lambda)$. For the four-port optical device platform to practically satisfy the requirements for optical summation, inversion and small-signal amplification functions, a realistic material system is required, in which the balance among absorption, reflection and transmission can be engineered at will. We thus consider planar photonic metamaterials—artificial electromagnetic media that are periodically structured on the subwavelength scale—as ultrathin absorbing films. We first demonstrate that the scattering matrix of Equation (4) can reasonably approximate the characteristics of realistic plasmonic metamaterial thin films and that the optical properties required to realize four-port coherent control devices are attainable in such media. Metamaterials fabricated from noble metals, conductive oxides and nitrides, graphene, topological insulators and dielectrics can exhibit strong dispersion, which may be employed to tailor the required optical response. We consider a generic photonic metamaterial structure that has been used and optimized in several previous studies,[7,46] including recent experimental demonstrations of coherent absorption modulation[34,35]—an array of asymmetric split ring (ASR) slits fabricated in a thin plasmonic metal film (specifically, after Refs. 34 and 35, a 430 nm period array in 50 nm of gold supported on a 30 nm silicon nitride membrane), as illustrated in Figure 3a.

The transmission, absorption and reflection characteristics of this type of metamaterial are well known and can be numerically calculated using a three-dimensional Maxwell solver, as illustrated in Figures 3b and 3c. They characteristically present a well-defined plasmonic absorption resonance at a wavelength prescribed by the material composition and unit cell geometry. The model system obviously has a finite thickness and the reflection and absorption coefficients for the two propagation directions differ slightly due to the bilayer composition of the metamaterial, with metal on one side and dielectric on the other (the transmission coefficients are identical, as they must be in a linear reciprocal system). Nonetheless, the metamaterial can reasonably be described by a simplified matrix that contains only one wavelength-dependent complex scattering parameter $s_M(\lambda)$, which is defined such that it assigns equal weight to each of the $S_{ij}$ coefficients of the real structure (detailed in Supplementary Information):

$$S_M(\lambda) = \frac{1}{4}[S_{11} + (S_{12}-1) + (S_{21}-1) + S_{22}] \\ = \frac{1}{4}[S_{11} + S_{12} + S_{21} + S_{22} - 2] \quad (6)$$

Transmission, reflection and absorption spectra calculated using $s_M(\lambda)$ are presented in Figure 3d and show good agreement with the spectra obtained *via* three-dimensional finite element modeling (Figures 3b and 3c).

A four-port device that operates as an optical amplifier can be considered analogous to a bipolar junction transistor common-emitter amplifier or its field effect transistor (FET) analog—the common-gate amplifier, with optical intensities $I_\beta$ and $I_\gamma$ representing signal voltages $V_{in}$ and $V_{out}$ respectively (Figure 4). The optical device can be considered to replicate transistor functionality if it operates in the small-signal amplification regime, whereby (for constant $I_\alpha$) the output signal intensity change $\Delta I_\gamma$ is greater than the input signal change $\Delta I_\beta$, i.e., a mode in which small intensity modulations at a given input translate to large amplitude modulations at a given output (the four-port device cannot be employed as a DC (direct current) amplifier because its output level is biased.) The small signal gain is defined as $G=\Delta I_\gamma/\Delta I_\beta$.

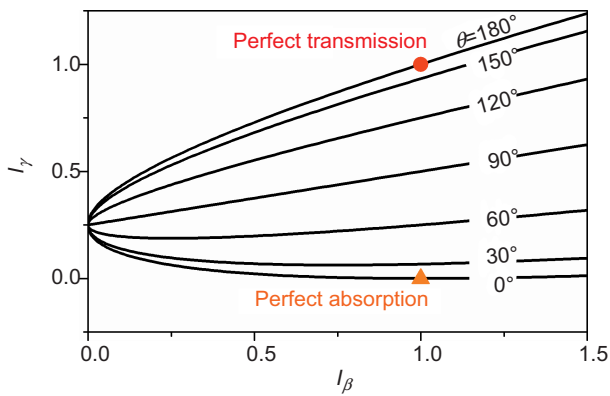

**Figure 2** Nonlinear character of a generic four-port device. Dependence of output intensity $I_\gamma$ on input intensity $I_\beta$ for a fixed input intensity $I_\alpha = 1$ and different phase retardations between $E_\alpha$ and $E_\beta$ ($s(\lambda) = -0.5$).

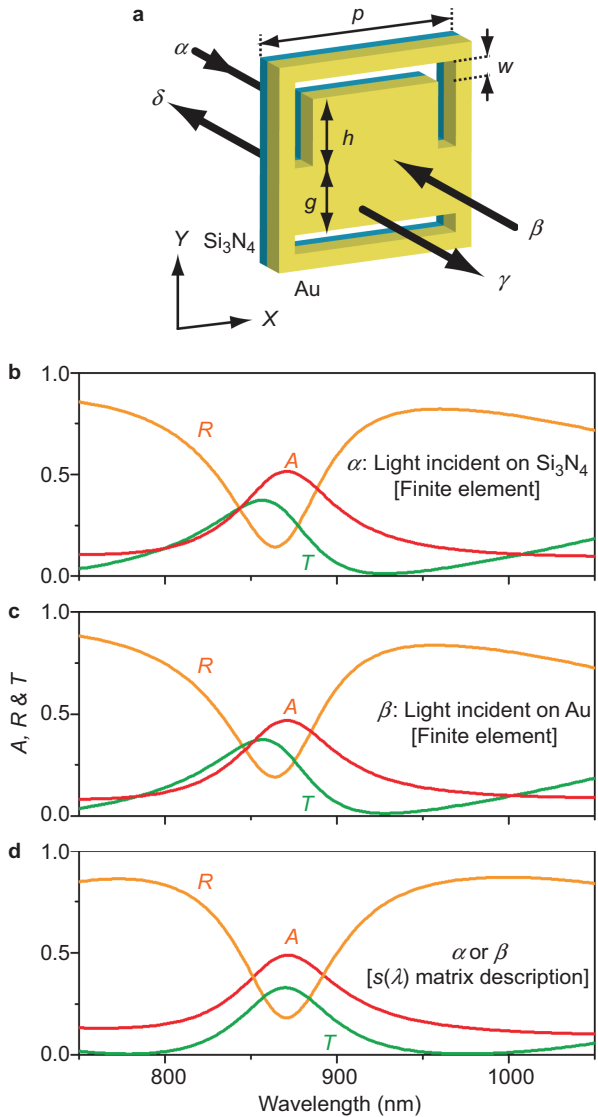

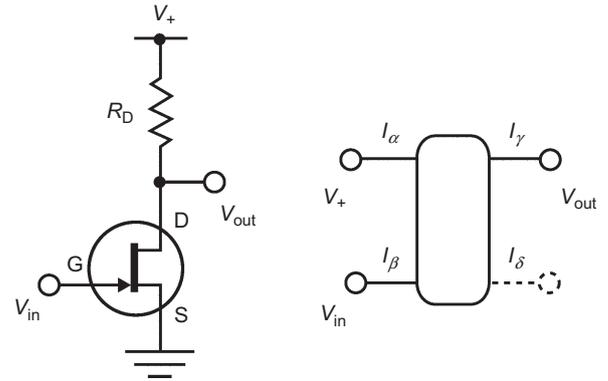

**Figure 3** Comparison of three-dimensional numerical and simplified matrix models for an asymmetric split ring (ASR) metamaterial. (**a**) Au-on-Si$_3$N$_4$ metamaterial unit cell geometry used in finite element numerical simulations, which employ a Drude–Lorentz model for gold, assign a fixed refractive index of 2.0 to silicon nitride (zero losses), and utilize periodic boundary conditions in the $x$ and $y$ directions (i.e., assume an infinite planar metamaterial array). Unit cell size $p=430$ nm; linewidth $w=40$ nm; arm and gap sizes $h$, $g=140$ nm. (**b**, **c**) Computationally modeled single-beam absorption $A$, reflection $R$ and transmission $T$ spectra for waves $\alpha$ and $\beta$, which are normally incident respectively on the dielectric membrane side and the metal film side of a gold/silicon nitride ASR metamaterial. (**d**) $A$, $R$ and $T$ spectra calculated using a simplified, propagation direction-independent single-beam scattering matrix $s_M(\lambda)$ (Equation (6)) to describe the properties of the metamaterial.

**Figure 4** Optical amplifier: a four-port coherent device (right) can be configured to operate in a regime analogous to a field effect transistor (FET) amplifier (left).

Amplifier functionality can be analyzed on the basis of Equation (5), from which the differential gain is

$$G = \frac{dI_\gamma}{dI_\beta} = |s(\lambda)|^2 + \text{Re}\{(1+s(\lambda))s(\lambda)^* e^{i\theta}\}\sqrt{\frac{I_\alpha}{I_\beta}} \quad (7)$$

In principle, $G$ can become infinitely large as $I_\beta$ approaches zero. Figure 5a illustrates, on the complex $s(\lambda)$ plane, the level of optical gain that can be achieved using the exemplar ASR metamaterial. The circle $P$ is the zero-loss contour for functional material layers; i.e., it encloses an area occupied by lossy media such as the metamaterial absorber, which is described by the line $M$ (Equation (6)). The gain contours $G$ are given by Equation (7) with an input intensity ratio $I_\alpha/I_\beta=100$ and a mutual phase $\theta=3\pi/4$ between the incident waves in the metamaterial plane. Differential gain is a function of wavelength and it attains a peak value of approximately 3.4 at $\lambda=846$ nm under these conditions, as shown by the track of line $M$ across the $G$ contours and by Figure 5b, which compares spectral dependencies derived from the single $s_M(\lambda)$ scattering parameter of Equation (6) and from the $S_{ij}$ coefficients of the metamaterial (i.e., taking into account the directional asymmetry illustrated in Figure 3b and 3c).

Figure 6 further details the performance of the four-port metamaterial device as an optical amplifier. Figure 6a shows output intensity $I_\gamma$ against input intensity $I_\beta$, as in Figure 2, for the peak-gain wavelength $\lambda=846$ nm given by the $s_M(\lambda)$ scattering parameter in Figure 5. For a phase retardation $\theta=3\pi/4$ between $E_\alpha$ and $E_\beta$ at the metamaterial plane, gain $>1$ can be observed if the working point $I_0$ is set below $I_\beta=0.3$ ($I_\alpha=1$). The phase response of the amplifier (Figure 6b), which is defined as the phase difference $\varphi$ between the carriers of the input and output signals, is nearly flat at a value $\varphi\approx 165°$ for the $\theta=3\pi/4$ input retardation setting, indicating that the amplified output will be in near-antiphase with the input signal.

The nonlinear response of the four-port device can also be employed to achieve optical summation and inversion. A summator or analog AND gate is configured with two input ports ($\alpha$ and $\beta$) and one output port ($\gamma$). Its binary logic truth table should be as follows:

$$\alpha = 1, \beta = 1 \rightarrow \gamma = 1$$

$$\alpha = 0, \beta = 1 \rightarrow \gamma = 0$$

$$\alpha = 1, \beta = 0 \rightarrow \gamma = 0$$

$$\alpha = 0, \beta = 0 \rightarrow \gamma = 0$$

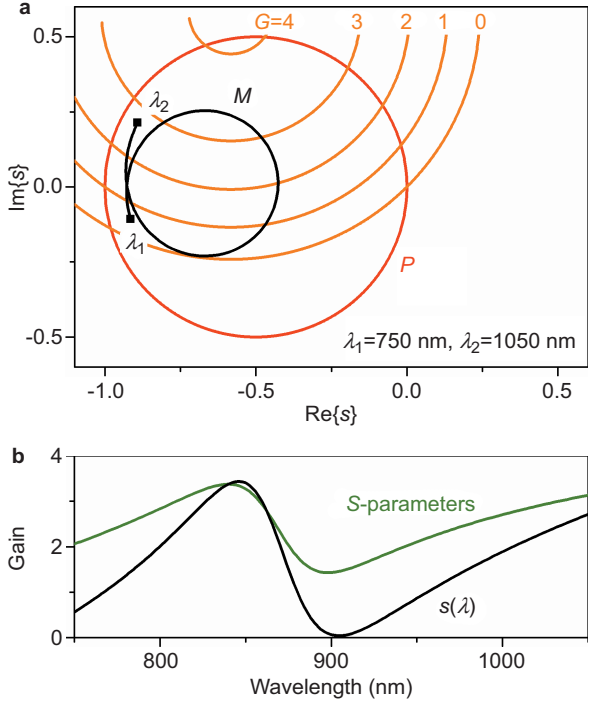

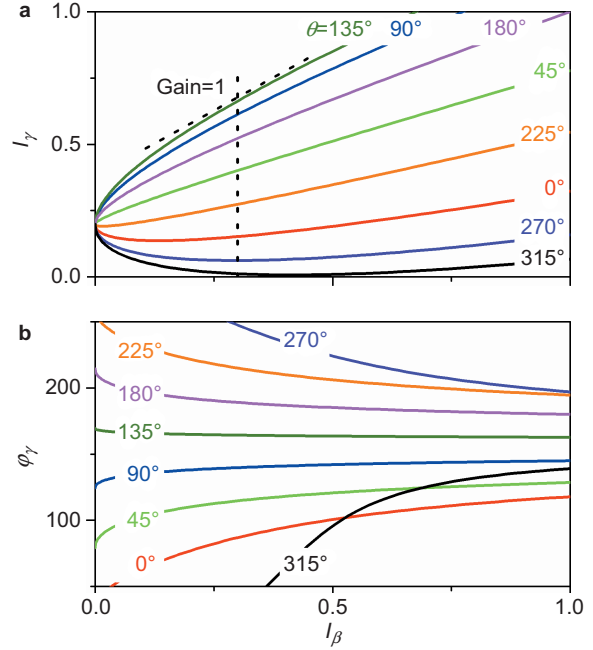

Figure 5 Conditions for differential gain in the four-port device. (a) In the complex $s(\lambda)$ plane, line $P$ encloses lossy media including the gold/silicon nitride metamaterial, which has properties described by line $M$ (N.B. wavelength is not directly proportional to distance along the line). These are overlaid with differential gain contours denoting values of $G$ between 0 and 4 for an input intensity ratio $I_\alpha/I_\beta=100$ and mutual phase $\theta=3\pi/4$. (b) Gain $G$ as a function of wavelength; the two curves are derived from (in black) the approximate $s_M(\lambda)$ scattering parameter of Equation (6) and (in green) the $S_{ij}$ coefficients of the metamaterial.

Figure 6 Performance of the four-port device as an optical amplifier. Amplifier throughput characteristics, (a) output intensity $I_\gamma$ and (b) output phase $\varphi_\gamma$, as functions of input intensity $I_\beta$ ($I_\alpha=1$) at a wavelength of 846 nm (the $s_M(\lambda)$ gain peak of Figure 5). A family of curves is presented in each panel for different values of the input phase retardation $\theta$ between $E_\alpha$ and $E_\beta$ (between 0° and 315° in steps of 45°, as labeled).

Optically, the output logic levels can be defined in terms of the output wave intensity $I_{out}=E_\gamma E_\gamma^*$ with respect to the input wave intensities $E_\alpha E_\alpha^*$ and $E_\beta E_\beta^*$ (each set to zero or $I_{in}$). Binary levels may be defined such that if $I_{out}<\mu I_{in}$ (where $\mu$ may be set between 0 and 2), then the output state of the AND gate is considered to be a logical '0'. If the output intensity $I_{out}>\mu I_{in}$, then the output state of the gate will be considered to be a logical '1'. Energy conservation requirements, the above truth table and the matrix Equation (4) then together prescribe that the functionality of a summator can be realized when the following conditions on $s(\lambda)$ are simultaneously satisfied:

$$|1+s(\lambda)|^2 + |s(\lambda)|^2 \leq 1 \qquad (8:P)$$

$$|(1+s(\lambda))e^{i\theta} + s(\lambda)|^2 > \mu \qquad (8:SUM1)$$

$$|s(\lambda)|^2 < \mu \qquad (8:SUM2)$$

$$|1+s(\lambda)|^2 < \mu \qquad (8:SUM3)$$

The first requirement ($P$) dictates that the output power cannot exceed the total input power for a functional layer without gain, i.e., in the diagrammatic representation of the complex $s(\lambda)$ plane of Figure 7, physical solutions lie on or inside the zero-loss $P$ contour. The remaining three conditions (SUM1, SUM2 and SUM3) define a set of intersecting inclusive (SUM2 and SUM3) and exclusive (SUM1) circular domains that delineate a parameter space within which thin-film media can facilitate optical summation. As shown in Figure 7, this space can be accessed within a certain wavelength range using the metamaterial structure described by $s_M(\lambda)$—the limits on this range being functions of the metamaterial design, the logic discrimination level $\mu$ and the input phase $\theta$.

An invertor or optical NOT gate is configured with one signal input port ($\alpha$) and one signal output ($\gamma$) with the following truth table:

$$\alpha = 1 \leftrightarrow \gamma = 0$$

$$\alpha = 0 \leftrightarrow \gamma = 1$$

In the four-port coherent optical implementation, a second 'bias' input ($\beta$) is also required. Again, output logic levels can be defined in terms of the output wave intensity $I_{out}=E_\gamma E_\gamma^*$ with respect to the input intensity $I_{in}$ (the signal input $E_\alpha E_\alpha^*$ may be set to zero or $I_{in}$, while the bias $E_\beta E_\beta^*$ is fixed at $I_{in}$) using the $\mu$ parameter. $s(\lambda)$ conditions for the realization of invertor functionality are thus:

$$|1+s(\lambda)|^2 + |s(\lambda)|^2 \leq 1 \qquad (9:P)$$

$$|(1+s(\lambda))e^{i\theta} + s(\lambda)|^2 < \mu \qquad (9:INV1)$$

$$|s(\lambda)|^2 > \mu \qquad (9:INV2)$$

The required material parameters can again be achieved using the metamaterial structure described by $s_M(\lambda)$, albeit in a different spectral range from the summator, as illustrated in Figure 8.

## CONCLUSIONS

We have illustrated that a four-port optical device based on an ultrathin (substantially subwavelength) material film can be configured as a small-signal optical amplifier, a summator and an invertor, and that planar photonic metamaterials can realistically provide the combination of properties necessary to deliver these functionalities. It is shown that a single generic asymmetric split ring metamaterial design can satisfy the requirements of all three functions in certain wavelength bands. The freedom with which the resonant dispersion of reflection, transmission and absorption can be manipulated by design in a metamaterial (ASR or otherwise) enables these operational ranges to be set at any desired wavelength. Moreover, metamaterial-based four-port optical devices may be dynamically tuned or switched by engaging nano-mechanically reconfigurable planar nanostructures[8,47] or metamaterials that are hybridized with optically/electrically/thermally activated functional media (e.g., phase-change materials,[9,10] liquid crystals[48,49] and semiconductors[50–52]).

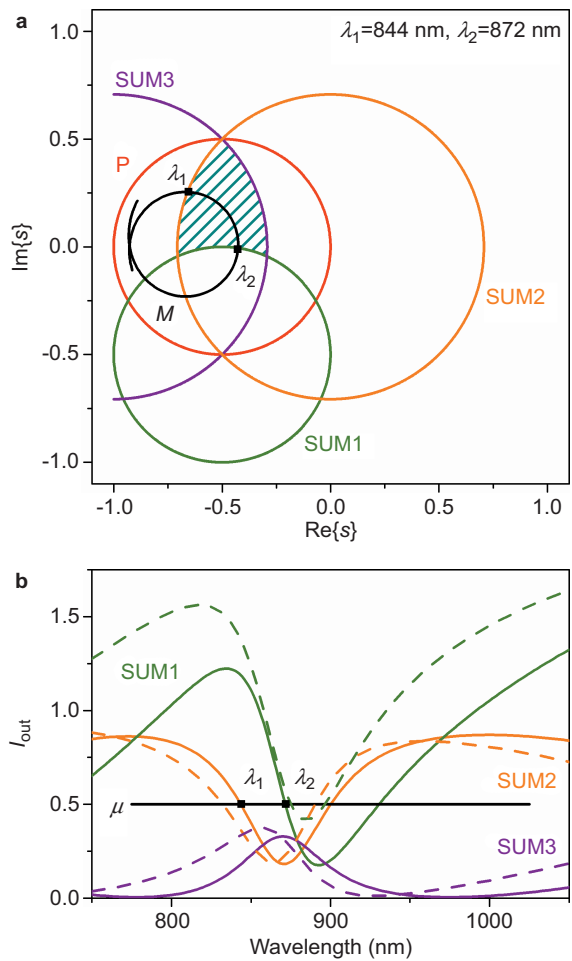
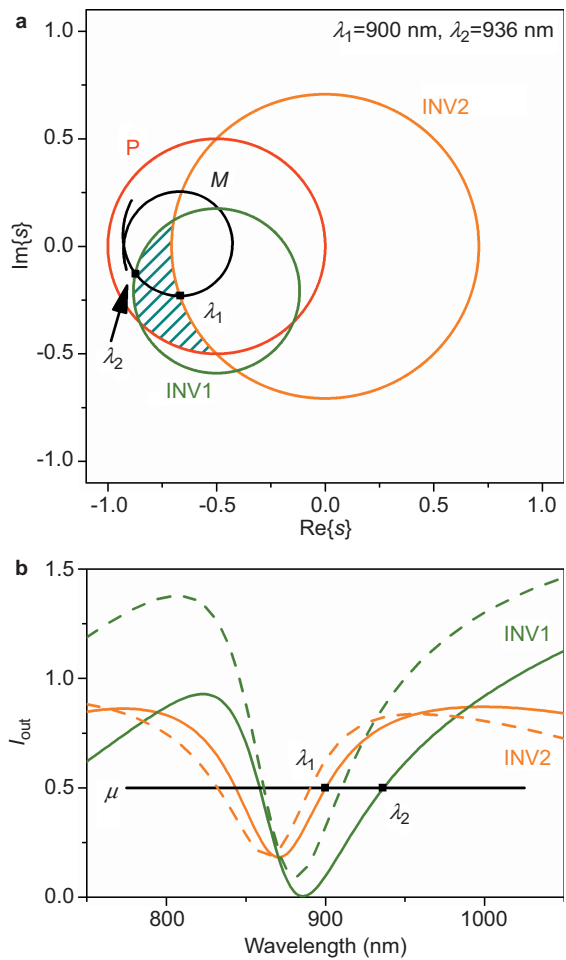

**Figure 7** Four-port optical device as a summator. (**a**) The hatched area in the complex $s(\lambda)$ plane, inside contours P, SUM2 and SUM3 and outside SUM1 (presented here for $\theta=\pi/2$ and $\mu=0.5$), represent the parameter space within which the four-port optical device can function as a summator. Under these conditions, the metamaterial described by line $M$ can access this domain between $\lambda_1=844$ and $\lambda_2=872$ nm. (**b**) Spectral dispersion of the SUM1, 2 and 3 conditions on $I_{out}$ (again for $\theta=\pi/2$). The solid lines are derived from the $s_M(\lambda)$ approximation; the dashed lines are derived from the $S_{ij}$ coefficients of the metamaterial. In this representation, the device can function when the logic discrimination level $\mu$ lies below SUM1 and above SUM2 and SUM3.

**Figure 8** Four-port optical device as an invertor. (**a**) The hatched area in the complex $s(\lambda)$ plane, inside contours P and INV1 and outside INV2 (presented here for $\theta=\pi/4$ and $\mu=0.5$), represents the parameter space within which the four-port optical device can function as an invertor. In these conditions, the metamaterial described by line $M$ can access this domain between $\lambda_1=900$ nm and $\lambda_2=936$ nm. (**b**) Spectral dispersion of the INV1 and 2 conditions on $I_{out}$ (again for $\theta=\pi/4$). The solid lines are derived from the $s_M(\lambda)$ approximation; the dashed lines are derived from the $S_{ij}$ coefficients of the metamaterial. In this representation the device can function when the logic discrimination level $\mu$ lies above INV1 and below INV2.

Such tunable four-port devices are relevant to the family of dissipation-less controllable crossover switches known as Fredkin gates—interconnected networks of which give rise to reversible non-Boolean 'direct logic' data processing architectures.[14] Reconfigurable four-port metamaterial devices may also be useful in other cognitive network applications, such as photonic oracles—optical networks mapped to represent NP-hard complexity decision problems, and dynamic networks configured as matrix inversion calculators.[16,17]

Networks that rely on locally coherent information carriers are being intensely investigated as part of the photonic telecommunications roadmap because they offer increased bandwidth *via* access to additional degrees of freedom including the phase and polarization of light, and to a variety of spectrally efficient modulation formats.[53] Four-port metamaterial logical devices could perform useful signal processing and routing functions in such networks.

Coherent gates hold significant advantage over nonlinearity-based data processing devices in their multi-THz bandwidth,[34–36] their freedom from harmonic distortion and their ability to operate at the single quantum level.[37] However, because the devices depend fundamentally on the interference of light waves, this comes at the expense of a requirement for precise positional settings to maintain relative phases, though this issue that may be addressed in a monolithic (e.g., silicon-photonic) platform that minimizes sources of differential movement among optical circuit components. Among the behavioral characteristics required of any all-optical data processing platform,[3] cascadability is also non-trivial in relation to such devices as complex assemblies require that the output of one element serves as the input to the next. Nonetheless, certain signal routing functions, or applications beyond data networks in sensing or spectroscopy, may be well served by singular optical logic or small-signal amplification elements.

## ACKNOWLEDGEMENTS

This study was supported by the Engineering and Physical Sciences Research Council (grant EP/G060363/1), the Royal Society, and the Singapore Ministry of Education [Grant MOE2011-T3-1-005]. Following a period of embargo, the data from this paper can be obtained from the University of Southampton ePrints research repository, DOI: 10.5258/SOTON/376804.

1. Richardson DJ. Filling the light pipe. *Science* 2010; **330**: 327–328.
2. Miller DAB. Device requirements for optical interconnects to silicon chips. *Proc IEEE* 2009; **97**: 1166–1185.
3. Miller DAB. Are optical transistors the logical next step? *Nat Photonics* 2010; **4**: 3–5.
4. Sharfin WF, Dagenais M. Femtojoule optical switching in nonlinear semiconductor laser amplifiers. *Appl Phys Lett* 1986; **48**: 321–322.
5. Sánchez M, Wen PY, Gross M, Esener S. Nonlinear gain in vertical-cavity semiconductor optical amplifiers. *IEEE Photonics Technol Lett* 2003; **15**: 507–509.
6. Nozaki K, Tanabe T, Shinya A, Matsuo S, Sato T et al. Sub-femtojoule all-optical switching using a photonic-crystal nanocavity. *Nat Photonics* 2010; **4**: 477–483.
7. Zheludev NI, Kivshar YS. From metamaterials to metadevices. *Nat Mater* 2012; **11**: 917–924.
8. Zhang JF, MacDonald KF, Zheludev NI. Nonlinear dielectric optomechanical metamaterials. *Light Sci Appl* 2013; **2**: e96; doi:10.1038/lsa.2013.52.
9. Gholipour B, Zhang JF, MacDonald KF, Hewak DW, Zheludev NI. An all-optical, non-volatile, bidirectional, phase-change meta-switch. *Adv Mater* 2013; **25**: 3050–3054.
10. Driscoll T, Kim HT, Chae BG, Kim BJ, Lee YW et al. Memory metamaterials. *Science* 2009; **325**: 1518–1521.
11. Caulfield HJ, Dolev S. Why future supercomputing requires optics. *Nat Photonics* 2010; **4**: 261–263.
12. Tucker RS. The role of optics in computing. *Nat Photonics* 2010; **4**: 405.
13. Miller DAB. The role of optics in computing. *Nat Photonics* 2010; **4**: 406.
14. Hardy J, Shamir J. Optics inspired logic architecture. *Opt Express* 2007; **15**: 150–165.
15. Cohen E, Dolev S, Frenkel S, Kryzhanovsky B, Palagushkin A et al. Optical solver of combinatorial problems: nanotechnological approach. *J Opt Soc Am A* 2013; **30**: 1845–1853.
16. Wu K, García de Abajo FJ, Soci C, Shum PP, Zheludev NI. An optical fiber network oracle for NP-complete problems. *Light Sci Appl* 2014; **3**: e147; doi:10.1038/lsa.2014.28.
17. Wu K, Soci C, Shum PP, Zheludev NI. Computing matrix inversion with optical networks. *Opt Express* 2014; **22**: 295–304.
18. Shapiro M, Brumer PW. *Principles of the Quantum Control of Molecular Processes*. New York: John Wiley & Sons Inc.; 2003.
19. Reiter DE, Sherman EY, Najmaie A, Sipe JE. Coherent control of electron propagation and capture in semiconductor heterostructures. *Europhys Lett* 2009; **88**: 67005.
20. Assion A, Baumert T, Bergt M, Brixner T, Kiefer B et al. Control of chemical reactions by feedback-optimized phase-shaped femtosecond laser pulses. *Science* 1998; **282**: 919–922.
21. Stockman MI, Faleev SV, Bergman DJ. Coherent control of femtosecond energy localization in nanosystems. *Phys Rev Lett* 2002; **88**: 067402.
22. Li XT, Stockman MI. Highly efficient spatiotemporal coherent control in nanoplasmonics on a nanometer-femtosecond scale by time reversal. *Phys Rev B* 2008; **77**: 195109.
23. Choi SB, Park DJ, Jeong YK, Yun YC, Jeong MS et al. Directional control of surface plasmon polariton waves propagating through an asymmetric Bragg resonator. *Appl Phys Lett* 2009; **94**: 063115.
24. Utikal T, Stockman MI, Heberle AP, Lippitz M, Giessen H. All-optical control of the ultrafast dynamics of a hybrid plasmonic system. *Phys Rev Lett* 2010; **104**: 113903.
25. Geiselmann M, Utikal T, Lippitz M, Giessen H. Tailoring the ultrafast dynamics of the magnetic mode in magnetic photonic crystals. *Phys Rev B* 2010; **81**: 235101.
26. Kao TS, Jenkins SD, Ruostekoski J, Zheludev NI. Coherent control of nanoscale light localization in metamaterial: creating and positioning isolated subwavelength energy hot spots. *Phys Rev Lett* 2011; **106**: 085501.
27. Gjonaj B, Aulbach J, Johnson PM, Mosk AP, Kuipers L et al. Active spatial control of plasmonic fields. *Nat Photonics* 2011; **5**: 360–363.
28. Wan WJ, Chong YD, Ge L, Noh H, Stone AD et al. Time-reversed lasing and interferometric control of absorption. *Science* 2011; **331**: 889–892.
29. Li ZP, Zhang SP, Halas NJ, Nordlander P, Xu HX. Coherent modulation of propagating plasmons in silver-nanowire-based structures. *Small* 2011; **7**: 593–596.
30. Miyata M, Takahara J. Excitation control of long-range surface plasmons by two incident beams. *Opt Express* 2012; **20**: 9493–9500.
31. Kao TS, Rogers ETF, Ou JY, Zheludev NI. ''Digitally'' addressable focusing of light into a subwavelength hot spot. *Nano Lett* 2012; **12**: 2728–2731.
32. Yoon JW, Koh GM, Song SH, Magnusson R. Measurement and modeling of a complete optical absorption and scattering by coherent surface plasmon-polariton excitation using a silver thin-film grating. *Phys Rev Lett* 2012; **109**: 257402.
33. Brinks D, Castro-Lopez M, Hildner R, van Hulst NF. Plasmonic antennas as design elements for coherent ultrafast nanophotonics. *Proc Natl Acad Sci USA* 2013; **110**: 18386–18390.
34. Zhang JF, MacDonald KF, Zheludev NI. Controlling light-with-light without nonlinearity. *Light Sci Appl* 2012; **1**: e18; doi:10.1038/lsa.2012.18.
35. Fang X, Tseng ML, Ou JY, MacDonald KF, Tsai DP et al. Ultrafast all-optical switching via coherent modulation of metamaterial absorption. *Appl Phys Lett* 2014; **104**: 141102.
36. Nalla V, Vezzoli S, Valente J, Sun H, Zheludev NI. 100 THz optical switching with plasmonic metamaterial. CLEO/Europe-EQEC 2015; Paper CF-9.6.
37. Roger T, Heitz J, Westerberg N, Gariepy G, Bolduc E et al. Coherent perfect absorption in the single photon regime. *Nat Commun* 2015; **6**: 7031.
38. Mousavi SA, Plum E, Shi JH, Zheludev NI. Coherent control of birefringence and optical activity. *Appl Phys Lett* 2014; **105**: 011906.
39. Shi JH, Fang X, Rogers ETF, Plum E, MacDonald KF et al. Coherent control of Snell's law at metasurfaces. *Opt Express* 2014; **22**: 21051–21060.
40. Pozar DM. *Microwave Engineering*. 2nd ed. New York: John Wiley & Sons Inc.; 1998. p196.
41. Jackson JD. *Classical Electrodynamics*. 3rd ed. New York: John Wiley & Sons; 1998. p501.
42. Hägglund C, Apell SP, Kasemo B. Maximized optical absorption in ultrathin films and its application to plasmon-based two-dimensional photovoltaics. *Nano Lett* 2010; **10**: 3135–3141.
43. Thongrattanasiri S, Koppens FHL, García de Abajo FJ. Complete optical absorption in periodically patterned graphene. *Phys Rev Lett* 2012; **108**: 047401.
44. Dutta-Gupta S, Martin OJF, Dutta Gupta S, Agarwal GS. Controllable coherent perfect absorption in a composite film. *Opt Express* 2012; **20**: 1330–1336.
45. Pu MB, Feng Q, Wang M, Hu CG, Huang C et al. Ultrathin broadband nearly perfect absorber with symmetrical coherent illumination. *Opt Express* 2012; **20**: 2246–2254.
46. Plum E, Tanaka K, Chen WT, Fedotov VA, Tsai DP et al. A combinatorial approach to metamaterials discovery. *J Opt* 2011; **13**: 055102.
47. Ou JY, Plum E, Zhang JF, Zheludev NI. An electromechanically reconfigurable plasmonic metamaterial operating in the near-infrared. *Nat Nanotechnol* 2013; **8**: 252–255.
48. Kang BY, Woo JH, Choi E, Lee HH, Kim ES et al. Optical switching of near infrared light transmission in metamaterial-liquid crystal cell structure. *Opt Express* 2010; **18**: 16492–16498.
49. Buchnev O, Ou JY, Kaczmarek M, Zheludev NI, Fedotov VA. Electro-optical control in a plasmonic metamaterial hybridised with a liquid-crystal cell. *Opt Express* 2013; **21**: 1633–1638.
50. Dani KM, Ku Z, Upadhya PC, Prasankumar RP, Brueck SRJ et al. Subpicosecond optical switching with a negative index metamaterial. *Nano Lett* 2009; **9**: 3565–3569.
51. Cho DJ, Wu W, Ponizovskaya E, Chaturvedi P, Bratkovsky AM et al. Ultrafast modulation of optical metamaterials. *Opt Express* 2009; **17**: 17652–17657.
52. Jun YC, Gonzales E, Reno JL, Shaner EA, Gabbay A et al. Active tuning of mid-infrared metamaterials by electrical control of carrier densities. *Opt Express* 2012; **20**: 1903–1911.
53. Kikuchi K. Coherent optical communications: historical perspectives and future directions. In: Nakazawa M, Kikuchi K, Miyazaki T, editors. *High Spectral Density Optical Communication Technologies*. Vol. 6. Optical and Fiber Communications Reports. Chapter 2. Berlin/Heidelberg: Springer-Verlag; 2010. p11–49.

Supplementary information for this article can be found on the *Light: Science & Applications*' website (http://www.nature.com/lsa/).